\documentclass[11pt,twoside]{article}  
\usepackage{apn3conf}

\newcommand{\degree}{$^\circ$}

\begin{document}   

%
%
%
%

\title{The Mid-IR Emission Structure of IRAS 16594-4656}
\titlemark{Mid-IR Structure of IRAS 16594-4656}

%
%
%

\author{Domingo An\'{\i}bal Garc\'{\i}a-Hern\'andez, Arturo Manchado\altaffilmark{1}}
\affil{IAC, 38205 La Laguna, 
Tenerife, Spain}
\author{Pedro Garc\'{\i}a-Lario}
\affil{ISO Data Centre, ESA, VILSPA, 28080 Madrid, Spain}
\author{Antonio Ben\'{\i}tez Ca\~{n}ete}
\affil{Astr. Dpt., La Laguna University, 38205 La Laguna, Tenerife, Spain}
\author{Jos\'e A. Acosta-Pulido, Ana M. P\'erez Garc\'{\i}a}
\affil{IAC, 38205 La Laguna, Tenerife, Spain}
\altaffiltext{1}{Consejo Superior de Investigaciones Cient\'{\i}ficas, CSIC}

%
%

\contact{D. A. Garc\'{\i}a-Hern\'andez}
\email{agarcia@ll.iac.es}

%
%
%
%
%

\paindex{Garc\'{\i}a-Hern\'andez, D. A.}
\aindex{Manchado, A.}     
\aindex{Garc\'{\i}a-Lario, P.}
\aindex{Ben\'{\i}tez Ca\~{n}ete, A.}
\aindex{Acosta-Pulido, J. A.}
\aindex{P\'erez Garc\'{\i}a, A. M.}

%
%

\authormark{Garc\'{\i}a-Hern\'andez et al.}

%
%



\begin{abstract}          
We report TIMMI2 diffraction-limited mid-IR images of the multipolar PPN
IRAS 16594-4656. By using the Lucy-Richardson deconvolution algorithm we
recover a two-peaked morphology in the innermost region at 8.6 $\mu$m and 11.5
$\mu$m. We interpret the observed mid-IR structure as the detection of the two
limb-brightened peaks indicating the presence of a dusty toroidal structure in
IRAS 16594-4656. We find that the supposed biconical openings of the dust torus
are in good agreement with one of the bipolar outflows identified in the HST
optical images.
\end{abstract}

%
%

\section{IRAS 16594-4656}

IRAS 16594-4656 (GLMP 507; hereafter I16594) was identified as a 
proto-planetary nebula (PPN) candidate on the basis of its IRAS colours by Volk
\& Kwok (1989). It shows a double-peaked spectral energy distribution which is
dominated by a strong mid- to far- infrared peak caused by dust emission that
is much brighter than the peak in the near-infrared. The HST optical images
show the presence of a bright central star surrounded by a multiple-axis
bipolar nebulosity (dominated by scattered light) with a complex morphology at
some intermediate viewing angle (Hrivnak, Kwok \& Su 1999).

There are several proofs of the presence of a circumstellar disc or torus (as
an equatorial density enhancement) in I16594. The highly collimated structure
seen in the HST optical images and the non-detected radio-continuum emission by
van de Steene \& Pottasch (1993) indicate that the emission lines observed in
the optical spectrum are the result of the shock excitation produced by a fast
bipolar wind from the central source. In agreement with this hypothesis
Garc\'{\i}a-Hern\'andez et al. (2002) detected H$_2$ shock-excited emission in
the direction of I16594. In addition, Su et al. (2003) observed 10\%
polarization around the central source, suggesting the presence of a
circumstellar torus. 

All previous studies of I16594 confirm the basic model of PPNe where an
enhancement of the equatorial density (torus) could obscure the central star in
visible light and is collimating the fast wind proceeding from the central
star. This model can be tested by obtaining high-resolution mid-IR images
of I16594. van de Steene, van Hoof, \& Wood (2000) did not
detect extended emission in their N-band images of I16594, but they observed
with a pixel scale of 0.66$\arcsec$", the weather conditions were very poor and
they did not correct for the PSF effects.

\section{Mid-IR observations and data reduction}
The observations were carried out in 2001 October 9 with the TIMMI2 mid-IR
camera coupled to ESO 3.6m telescope (La Silla). TIMMI2 has an array of 320 x
240 pixels with a pixel scale of 0.2$\arcsec$ x  0.2$\arcsec$. The
observational conditions were very good and we obtained the mid-IR images (at
8.6 $\mu$m[N1] and 11.5 $\mu$m[N11.9]) at the diffraction limit of the
telescope. We used the standard nodding/chopping mid-IR observational technique
and the basic data reduction process to obtain a final image per filter. The
flux calibration was made using the averaged conversion factors derived from
the observations of a number of standard stars at different air masses. We
removed the effects of the PSF by using the Lucy-Richardson deconvolution
algorithm implemented in IRAF in order to recover the emission structure for
each filter. In Figure 1 we display the original/raw images with the
corresponding deconvolved ones for every filter.

\section{Raw and deconvolved mid-IR morphology of the nebula}
We can identify two main mid-IR emission regions in I16594 in the two filters
(see Figure 1, left). An extended halo which is approximately elliptical
($\leq$40\% of the peak intensity) and an elongated emission core ($>$40\% of
the peak intensity). We fitted the emission halo with elliptical isophotes in
order to determine the source centre. This was determined by averaging the
central coordinates of the isophotes with 20$-$40\% of the peak intensity to
avoid any contamination from the core structure. The images displayed in Figure
1 have been centred on this position. 
Because the central star has a B7 spectral type (van de Steene, Wood, \& van
Hoof 2000), no contribution in 10 $\mu$m to the flux observed is expected. In
addition, Hrivnak, Kwok, \& Su (1999) found that the contribution of the
photospheric component to the total flux in the mid-IR is only $\sim$ 3\%. For
this reason, we have considered the photospheric contribution negligible.

The deconvolved N1 and N11.9 images are shown in Figure 1 on the right. We
recover a two-peaked morphology in the innermost region in the two filters. The
two detected peaks are approximately orientated along the north-south direction
(P.A.$\sim$$-$10$-$170\degree) and we can define a symmetry axis in the
east-western direction (P.A.$\sim$80$-$260\degree).  We interpret the mid-IR
morphology seen in the deconvolved images as the evidence of the two
limb-brightened peaks corresponding to a dusty toroidal structure in I16594. We
are sure that the I16594 deconvolved structure is real for two reasons: first
because we have a large signal-to-noise ratio in the raw images and second we
obtain a similar emission structure in N1 and N11.9. Otherwise, we find that
the north-peak (P.A.$\sim$$-$10\degree) is brighter and more extended than the
other. Ueta et al. (2001) found a similar asymmetric appearance in IRAS 22272+5435.
\begin{figure}
\epsscale{0.75}
\plotone{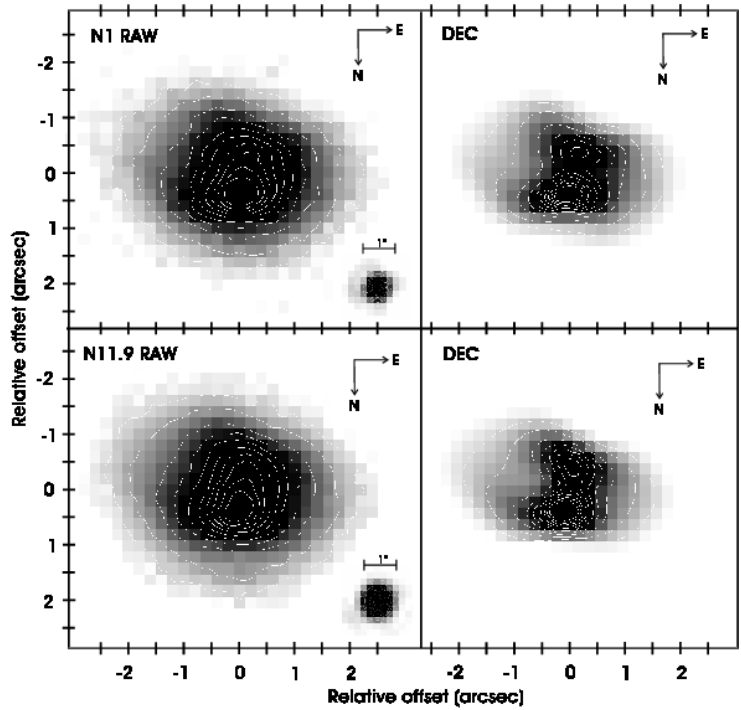}
\caption{Raw mid-IR images of I16594 in N1 (8.6 $\mu$m) and
N11.9 (11.5 $\mu$m) filters (left, from top to
bottom) and the corresponding deconvolved images (right). The thick marks show 
relative offsets from the centre of the nebula in arc seconds. 
Contours range from 10\% to 90\% of the peak intensity (in  
steps of 10\%) and we have added the outermost contours,  
which correspond to 5\% of the peak intensity. The insets show 
the standard star PSFs in each filter.\label{fig1}}
\end{figure}

\begin{figure}
\epsscale{0.75}
\plotone{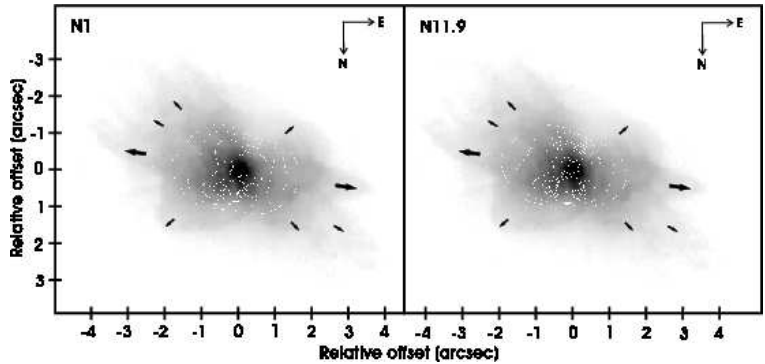}
\caption{Optical HST image of I16594 (taken from 
the HST Data Archive) overlaid with contours of the deconvolved 
mid-IR images in the N1 (8.6 $\mu$m) and N11.9 (11.5 $\mu$m) filters. 
The displaying conventions follow those of Fig.1. The big arrows indicate
the directions of the optical prominences that coincide with the supposed
biconical openings of the dust torus. The other bipolar axis directions 
identified in I16594 have been marked with small arrows.\label{fig2}}
\end{figure}

\section{Evidence for precession in IRAS 16594-4656}
We infer pairs of elongated structures with at least four different bipolar
axis at P.A. $\sim$34\degree, $\sim$54\degree, $\sim$84\degree and
$\sim$124\degree from the HST optical images. The presence of different bipolar
axis suggests that material has been ejected episodically from a
precessing/rotating source. Similar structures are also detected in more
evolved PNe (e.g. Guerrero \& Manchado 1998). We have displayed the I16594
optical HST image (taken from the HST Data Archive) overlaid with the contours
of the deconvolved mid-IR images in Figure 2. We find that the mid-IR emission
symmetry axis approximately coincides with the bipolar axis located at
P.A.$\sim$84\degree. The optical prominences towards the west
(P.A.$\sim$260\degree) and east (P.A.$\sim$80\degree) are in good agreement
with the supposed biconical openings of the dust torus. This finding may thus
indicate that the ejected  material proceeding from the central source is at
present collimated by the dust torus along the east-west direction
(P.A.$\sim$80$-$260\degree). The other bipolar outflows may be the result of
episodic mass loss processes and could be the signature of a precessing system.

\acknowledgments
PGL and AM acknowledges support from grant PB97-1435-C02-02 from the
Spanish Direcci\'on General de Ense\~{n}anza Superior (DGES).

%
%
%
%


\end{document}